\documentclass[prb, reprint]{revtex4-1}
\usepackage{hyperref}
\usepackage{geometry}                
\usepackage{graphicx}
\usepackage{natbib}
\usepackage{bm}
\usepackage{amssymb}
\usepackage{amsmath}
\usepackage{epstopdf}
\begin{document}

\title{Possible Nematic Order Driven by Magnetic Fluctuations in Iron Pnictides}
\author{Kok Wee \surname{Song}}
\author{Yung-Ching \surname{Liang}}
\author{Hokiat \surname{Lim}}
\author{Stephan \surname{Haas}}

\affiliation{Department of Physics and Astronomy, University of Southern California, California 90089 USA}

\date{\today}
\begin{abstract}
In this paper, instabilities of the isotropic metallic phase in iron pnictides are investigated. The relevant quartic fermionic interaction terms in the model are identified using phase space arguments. Using the functional integral formalism, a Hubbard-Stratonovich transformation is used to decouple these quartic terms. This procedure introduces several bosonic fields which describe the low-energy collective modes of the system. By studying the behavior of these collective modes, a possible instability is found in the forward scattering channel of the isotropic phase driven by magnetic fluctuations. Using mean field analysis, we obtain a static and homogeneous ground state. This ground state is metallic, but the electron Fermi pockets are distorted unequally at different pockets in momentum space. This results in a desirable nematic ordering which breaks the lattice $C_4$ symmetry but preserves translational symmetry and may explain several experimental observations.

\end{abstract}

\maketitle

\section{Introduction}

A new unconventional superconducting state has recently been discovered in doped iron pnictide (FeAs) materials .\cite{doi:10.1021/ja800073m,Stewart:2011fk,johnston2010puzzle} Although FeAs is similar to the cuprates in some respects, i.e. it is a layered and exhibits magnetic ordering in its parent compound, some of its electronic properties are fundamentally different. In contrast to cuprates, the electronic structure of FeAs is effectively described by a multi-band model.\cite{PhysRevB.75.035110,PhysRevLett.100.237003,PhysRevB.77.220506,1367-2630-11-2-025016} In addition, FeAs is believed to be less correlated than the cuprates. These observations suggest that their mechanism of superconductivity may be different. In FeAs, the Fermi surface (FS) breaks down into several pockets. Due to this feature, scattering processes at low energy can give rise to non-trivial many-body physics. This may explain the origin of superconductivity in FeAs with only repulsive interactions.\cite{doi:10.1146/annurev-conmatphys-020911-125055}

In addition to the superconducting state, another intriguing property in FeAs has recently been observed. The electronic properties exhibit a directional preference,\cite{Chu13082010,Chu10082012,0295-5075-93-3-37002,PhysRevB.81.184508,Chuang08012010,Nakajima26072011,Kasahara:2012fk} regardless of the system's intrinsic 4-fold crystal rotational symmetry ($C_4$ symmetry). This anisotropy may be related to nematicity due to the interactions between electrons.\cite{doi:10.1146/annurev-conmatphys-070909-103925} The new phase is metallic and associates with magnetic order. This observation has led to speculations about the relationship between magnetic fluctuations and nematic ordering. Some theoretical studies have shown that magnetic fluctuations can play an important role in giving rise to nematic order.\cite{PhysRevLett.101.076401,PhysRevB.77.224509,PhysRevB.78.020501,PhysRevB.85.024534} The functional renormalization group (FRG) point of view\cite{PhysRevB.80.064517,PhysRevLett.106.187003} provides an interesting perspective to understand the stability of the metallic phase in FeAs. Because of the intricate geometrical structure of the FS, the coupling parameters in the forward scattering channels are relevant under RG transformation instead of just being marginal. This suggests that some non-trivial scattering processes may occur in the direct channels which can potentially break the stability of the FS. 

Based on these observations, the central goal of this paper is to investigate how magnetic fluctuations can affect the direct channels in FeAs, and yield instabilities. We argue that nematic order is more favorable energetically than the isotropic metallic phase. This paper is organized as the following. In Sec. \ref{model}, we will discuss the effective model that will be used in the study. Sec. \ref{Instabilities} will show how the magnetic fluctuations affect the density fluctuations and give rise to instabilities in the isotropic metallic phase. Sec. \ref{MF}, the nematic order and its magnetic fluctuations will be discussed at the mean field level. In Sec. \ref{discussions}, we will briefly discuss these results in the context of experiments and other theoretical studies.

\section{Model Hamiltonian}\label{model}

\begin{figure}[htbp]
   \centering
   \includegraphics[scale=0.5]{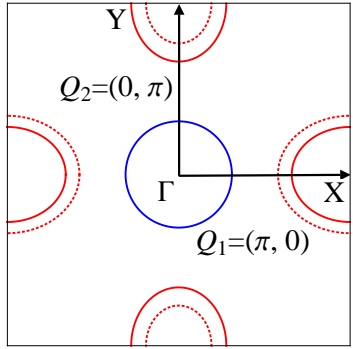} 
   \caption{A schematic diagram showing the FS structure of a typical FeAS in the unfolded Brillioun Zone: The bold curves represent the FS of the model in the isotropic metallic phase. the dotted curves represent the unequally renormalized FS pockets in the nematic phase with $\langle\Delta^-\rangle>0$.}
   \label{FSpocket}
\end{figure}

We extend the model of Ref. \onlinecite{PhysRevB.85.024534} by including the forward scattering (density-density) interactions within intra- and inter- FS pockets. The free part of the Hamiltonian is given by
\begin{equation}
\mathcal{H}_0=\sum_{\alpha,\mathbf{k}}\epsilon^\alpha_\mathbf{k}c^{\dagger}_{\alpha,\mathbf{k}s}c_{\alpha,\mathbf{k}s},
\end{equation}
where $s$ is the spin, and $\alpha$ represents the location of the FS pockets. $\alpha=\Gamma$ represents the hole FS pocket with its center located at $(0,0)$, and $\alpha=X,Y$ represent the electron FS pockets with centers located at $\mathbf{Q}_1=(\pi,0)$ and $\mathbf{Q}_2=(0,\pi)$ respectively (see Fig. \ref{FSpocket}). The energy dispersions are $\epsilon^{\Gamma}_{\mathbf{k}}=\epsilon_0-\frac{k^2}{2m}-\mu$, $\epsilon^{X}_{\mathbf{k}+\mathbf{Q}_1}=-\epsilon_0+\frac{k_x^2}{2m_x}+\frac{k_y^2}{2m_y}-\mu$, and $\epsilon^{Y}_{\mathbf{k}+\mathbf{Q}_2}=-\epsilon_0+\frac{k_x^2}{2m_y}+\frac{k_x^2}{2m_y}-\mu$, where $m$, $m_x$, and $m_y$ are the band masses, $\epsilon_0$ is the offset energy, and $\mu$ is the chemical potential. For simplicity, we shift the momentum of the $X$ ($Y$) pocket to $\mathbf{Q}_1$ ($\mathbf{Q}_2$) i.e. $\epsilon^X_{k+\mathbf{Q}_1}\rightarrow\epsilon^X_{k}$ ($\epsilon^Y_{k+\mathbf{Q}_2}\rightarrow\epsilon^Y_{k}$). Furthermore, summing over all spins is implicitly assumed throughout the paper.

In the low temperature limit, the dominant scattering processes only take place in the vicinity of the FS. Moreover, because of momentum conservation, all scattering processes are strictly constrained to the available phase space for two-body scattering processes. We are interested in the processes with small momentum transfer (or up to a nesting vector in exchange channels), as these processes are likely to develop collective modes.\cite{SimonQFT} Therefore, keeping only direct, exchange, and BCS channels in the model is sufficient, since these dominate over the others in phase space.  

Although most of the momentum channels are being discarded, there is still ambiguity which of these three special channels are relevant to the problem. We can further eliminate some of them by using the following intuitive arguments. First, for electron-hole (e-h) interactions, because of the FS pocket nesting spin-density-wave (SDW) fluctuations are the most important collective mode, especially as the system temperature approaches the magnetic ordering phase transition temperature ($T_N$). Therefore, the exchange channel in the e-h interactions is kept. Normally, the exchange channel of this interaction breaks into spin triplet and singlet channels, which correspond to SDW and charge-density-wave (CDW) respectively. However, the CDW fluctuations are insignificant and can be discarded, because their characteristic energy scale is much lower than for the SDW. By using similar arguments, all the direct and BCS channels in the e-h interactions can be discarded. For the electron-electron (e-e) interactions, since we wish to investigate the instabilities of the isotropic paramagnetic phase (metallic), the direct channels should be kept. Following the same reasoning for the e-h interactions case, the exchange and BCS channels are irrelevant in e-e interactions. Therefore, the effective interacting Hamiltonian becomes
\begin{equation}\label{Hint}
\begin{split}
\mathcal{H}_{\text{int}}=&-\frac{u_\text{spin}}{2}\sum_{\alpha=X,Y}\sum_{\mathbf{q}}\mathbf{s}_{\alpha,\mathbf{q}}\cdot\mathbf{s}_{\alpha,-\mathbf{q}}\\
&+\frac{u^{(1)}_4}{2}\sum_{\alpha=X,Y}\sum_{\mathbf{q}}\rho_{\alpha,\mathbf{q}}\rho_{\alpha,-\mathbf{q}}\\
&+u_6\sum_{\mathbf{q}}\rho_{X,\mathbf{q}}\rho_{Y,-\mathbf{q}},
\end{split}
\end{equation}
where $\mathbf{s}_{\alpha,\mathbf{q}}=\sum_{\mathbf{k}}c^{\dagger}_{\Gamma,\mathbf{k+q},s}\bm{\sigma}_{ss'}c_{\alpha,\mathbf{k}s'}$, $\bm{\sigma}$ is the Pauli matrix vector, and $\rho_{\alpha,\mathbf{q}}=\sum_\mathbf{k}c^{\dagger}_{\alpha,\mathbf{k+q},s}c_{\alpha,\mathbf{k}s}$. The first term of Eq. \eqref{Hint} combined with $\mathcal{H}_0$ is the original model given in Ref. \onlinecite{PhysRevB.85.024534}. $u_\text{spin}$ is the coupling constant for electron and hole pocket interactions in the triplet channels. The last two terms are newly introduced, describing the density-density interactions between the $X$ and $Y$ FS pockets. $u^{(1)}_4$ is the coupling constant for the intra-electron pocket interactions, and $u_6$ is the coupling constant for inter-electron pocket interactions. The notation of the coupling constants in \eqref{Hint} are the same as in Ref. \onlinecite{PhysRevB.82.214515}, and angular dependence has been ignored. Because of the $C_4$ symmetry in the system, the electron intra pocket interactions (second term in \eqref{Hint}) have the same coupling constants. The above interaction terms only capture dominant scattering processes of the electrons in the vicinity of the FS. The full interaction terms include arbitrary momentum transfer contributions as well.\cite{PhysRevB.82.214515} 

\section{Instabilities of the isotropic phase}\label{Instabilities}

To investigate whether the model in Sec. \ref{model} can develop any instability, one can use rigorous diagrammatic methods. However, we use a simple alternative approach to tackle the problem. This approach is based on the functional integration formalism, introducing auxiliary bosonic fields  to decouple the fermonic quartic terms (Hubbard-Stratonovich transformation). By integrating out the fermonic degrees of freedom, one can obtain the action of these bosonic fields which describes the collective modes of the system at low energy. In the following, we demonstrate this procedure and investigate the stabilities of the isotropic metallic phase.

\subsection{Hubbard-Stratonovich Transformation}

In order to perform the Hubbard-Stratonovich transformation, it is important to check whether $u^{(1)}_4-u_6$ is positive or negative. Moreover, we will see later that this is also the key parameter determining whether instabilities can emerge in the direct channels. For now, we assume $u^{(1)}_4-u_6$ is negative and set $u_6+u^{(1)}_4=u'_\text{n}$ and $u_6-u^{(1)}_4=u_\text{n}$ (positive). Then we introduce a set of auxiliary bosonic fields $\phi_q=(\Delta^+_q,\Delta^-_q, \mathbf{M}^X_q,\mathbf{M}^Y_q)$, where $q=(i\nu_n,\mathbf{q})$ and $\nu_n$ is the bosonic Matsubara frequency. 

By inserting the `fat' unity \eqref{Unity} into the partition function integrand ($\mathcal{Z}=\int\mathcal{D}[\bar{\Psi},\Psi]\mathbf{1}e^{-S}$), the fermonic quartic terms in the action are canceled out by shifting the auxiliary fields (\ref{shift1} - \ref{shift3}). By introducing the Nambu spinor $\bar{\Psi}_k=(\bar{\psi}^X_{k\uparrow},\bar{\psi}^X_{k\downarrow},\bar{\psi}^Y_{k\uparrow},\bar{\psi}^Y_{k\downarrow}, \bar{\psi}^\Gamma_{k\uparrow},\bar{\psi}^\Gamma_{k\downarrow})$, the new action can be compactly expressed as
\begin{equation}\label{S}
\begin{split}
S&=\sum_{kk'}\Bar{\Psi}_k\mathcal{G}_{kk'}^{-1}\Psi_{k'}+\frac{1}{u_\text{spin}}\sum_{\alpha,q}\mathbf{M}^\alpha_q\cdot \mathbf{M}^\alpha_{-q}\\
&+\sum_q\left[\frac{2}{u'_\text{n}}\Delta^+_q\Delta^+_{-q}+\frac{2}{u_\text{n}}\Delta^-_q\Delta^-_{-q}\right],
\end{split}
\end{equation}
where
\begin{equation}\label{propagator}
\mathcal{G}^{-1}_{kk'}=\mathcal{G}^{-1}_{0,kk'}+\mathcal{V}_{kk'}.
\end{equation}
$\mathcal{G}^{-1}_{0,kk'}$ is given by
\begin{equation}
\begin{pmatrix}G^{-1}_{X,k}& 0 & 0 \\ 0 &G^{-1}_{Y,k}& 0 \\
0 &  0 & G^{-1}_{\Gamma,k}\end{pmatrix}\delta_{kk'},
\end{equation}
and $\mathcal{V}_{kk'}$ is given by
\begin{equation}
\begin{pmatrix}i\Delta^+_{k-k'}+\Delta^-_{k-k'}& 0 &\mathbf{M}^X_{k-k'}\\
0&i\Delta^+_{k-k'}-\Delta^-_{k-k'}& \mathbf{M}^Y_{k-k'} \\
\mathbf{M}^X_{k-k'} &  \mathbf{M}^Y_{k-k'} & 0  \end{pmatrix},
\end{equation}
where $G^{-1}_{\alpha,k}=-i\omega_n+\epsilon^{\alpha}_{\mathbf{k}}$, and $\alpha=X,Y,\Gamma$. 

$\Delta^-$ must change sign under exchanging $x$ and $y$ axis to preserve the original $C_4$ symmetry in \eqref{S} because the boson-fermion coupling terms. Integrating out the fermionic fields, we obtain the action for the bosonic fields,
\begin{equation}\label{fullaction}
\begin{split}
S[\phi]&=-\text{Tr}\ln\mathcal{G}^{-1}+\frac{1}{2}\sum_{\alpha,q}\frac{1}{u_\text{spin}}M^\alpha_q M^\alpha_{-q}\\
&+\frac{1}{2}\sum_{q}\left(\frac{1}{u'_n}\Delta^+_q\Delta^+_{-q}+ \frac{1}{u_n}\Delta^-_{q}\Delta^-_{-q}\right).\\
\end{split}
\end{equation}
Note that $\mathbf{M}^{\alpha}=M^{\alpha}\mathbf{n}$, where $\mathbf{n}$ is some arbitrary direction of the spin. Since $\Delta^-$ change sign under $c_4$ rotation, any non-zero ground state expectation value of $\Delta^-$ implies spontaneous $C_4$ rotational symmetry breaking. Therefore, we can identify $\Delta^-$ as the nematic order parameter, and Eq. \eqref{fullaction} describes how the magnetic and nematic collective modes interact with each other.

\subsection{Isotropic metallic phase}
The extremum of the action is given by
\begin{equation}\label{Extremum}
\left.\frac{S[\phi]}{\delta\phi_q}\right|_{\phi_q=\langle\phi_q\rangle}=0.
\end{equation}
Solving this equation involves the inversion of $\mathcal{G}^{-1}$. However, without knowing the functional form of $\phi_q$, this is a formidable task, since $\mathcal{G}^{-1}$ is generally not diagonal in momentum-frequency space. Therefore, in order to proceed, we need to guess a solution which simultaneously satisfies $\sum_q\mathcal{G}_{kq}[\mathcal{G}^{-1}]_{qk'}=\mathbb{I}_{3\times3}\delta_{kk'}$ and Eq. \eqref{Extremum}. Since we want to investigate the stability of the metallic phase, one obvious and desirable choice is $\langle \mathbf{M}^X_q\rangle=\langle \mathbf{M}^Y_q\rangle=0$ (paramagnetic), and $\langle\Delta^-_q\rangle= 0$ (isotropic). In general $\langle\Delta^+_q\rangle$ can be a nonzero constant. For instance, \cite{PhysRevB.80.064517} the FS pockets can shrink or expand according to its non-zero expectation values. However, the hole and electron FS pockets must be renormalized, such that no net charge is introduced into the system (Luttinger theorem).\cite{PhysRev.119.1153} Modifying the FS pockets in this manner does not influence our result qualitatively, since angular dependence in the coupling constants is ignored. Therefore, we assume $\langle\Delta^+_q\rangle=0$.

Using the saddle point approximation, we expand Eq. \eqref{fullaction} around the extremum with $\langle\phi_q\rangle=0$ and include fluctuations beyond the mean field level. Thus, we obtain 
\begin{equation}\label{Expansion1}
\begin{split}
\text{tr}\ln\mathcal{G}^{-1}&=\text{tr}\ln[\mathcal{G}_0^{-1}(1+\mathcal{G}_0\mathcal{V})]\\
&=\text{tr}\ln\mathcal{G}_0^{-1}-\sum_n\frac{1}{2n}\text{tr}(\mathcal{G}_0\mathcal{V})^{2n},
\end{split}
\end{equation}
and
\begin{equation}
\mathcal{V}_{kk'}=\begin{pmatrix} \delta \Delta^-_{k-k'} & 0 & \delta M^X_{k-k'}
\\0& -\delta \Delta^-_{k-k'}  & \delta M^Y_{k-k'} \\
 \delta M^X_{k-k'} &  \delta M^Y_{k-k'} & 0 \end{pmatrix},
\end{equation}
where $\delta M^{X,Y}_{k-k'}$ and $\delta \Delta^-_{k-k'}$ are the fluctuations of the collective modes near $\langle\phi_q\rangle=0$. This yields the effective action
\begin{equation}\label{Seff}
\begin{split}
S_{\text{eff}}\simeq&\frac{1}{2}\sum_q\delta \Delta^-_{q}(\frac{2}{u_6-u^{(1)}_4}+2\Pi_q)\delta \Delta^-_{-q}\\
&+\frac{1}{2}\sum_{\alpha,q}\delta  M^\alpha_{q}\chi^{-1}_{\alpha,q}\delta M^\alpha_{-q}\\
&+\lambda\sum_{x,\alpha}(\delta \Delta^-_x)^2(\delta M^\alpha_x)^2,
\end{split}
\end{equation}
where $\Pi_q=\sum_kG_{\alpha,k}G_{\alpha,k+q}$, $\chi^{-1}_{\alpha,q}=\frac{2}{u_\text{spin}}+2\sum_kG_{\Gamma,k}G_{\alpha,k+q}$, and $\lambda=\sum_kG_{\Gamma,k}(G_{\alpha,k})^3$, and $x=(\tau, \mathbf{x})$ represent the imaginary time and real space coordinate respectively. Note that the frequency and momentum dependence in $\lambda$ can be discarded because of  power counting\cite{RevModPhys.84.299} as only local interactions are considered.\cite{PhysRevB.14.1165} Moreover, the other terms beyond Gaussian only yield higher order corrections for the action and are not important to the problem (before the magnetic order sets in), except in the last term of \eqref{Seff}, which accounts for the interactions between magnetic and nematic order fluctuations. The physical meaning of this term is that magnetic fluctuations associated with the fluctuations of the $X$ and $Y$ FS pockets unequally can lower the total energy, since this term is negative ($\lambda$ is negative). If the magnetic fluctuations are small, these contributions are negligible. When the magnetic fluctuations become large in the vicinity of the magnetic ordering transition point, the stability of the $\langle\Delta^-_q\rangle=0$ isotropic phase can break down.

\subsection{Instabilities due to magnetic fluctuations}

To find the instabilities, we can integrate out $\delta M^\alpha$. For small $q$, $\chi^{-1}_{\alpha,q}=r_0+\gamma|\nu_n|+f_{\alpha,\mathbf{q}}$, where $r_0=\chi^{-1}_{\alpha,q=0}$, $\gamma$ is the Landau damping coefficient, and $f_{\alpha,\mathbf{q}}=q_x^2(1\pm\eta)+q_y^2(1\mp\eta)+\eta_zq_z^2$ is a generalized anisotropic function with $-1<\eta<1$, and the upper (lower) sign refers to $\alpha=X$ ($\alpha=Y$).\cite{PhysRevB.85.024534} In this paper, we are not interested in the out-of-plane anisotropy. Therefore, we set $\eta_z$ to zero. We thus can obtain the effective action for $\delta \Delta^-$,
\begin{equation}
\begin{split}
S_{\text{eff}}&=\frac{1}{2}\sum_q\delta \Delta^-_q(\frac{2}{u_\text{n}}+2\Pi_q)\delta \Delta^-_{-q}\\
&+\text{tr}\ln[r_0+\gamma\partial_\tau+f_{\alpha,\hat{\mathbf{q}}}+\lambda(\delta \Delta^-_x)^2],
\end{split}
\end{equation}
where $\hat{\mathbf{q}}$ are momentum operators. The trace in the above equation can be evaluated in momentum-frequency space, yielding
\begin{equation}
\sum_{\alpha,q}\ln[r_0+\gamma|\nu_n|+f_{\alpha,\mathbf{q}}+\lambda\delta \Delta^-_q\delta \Delta^-_{-q}].
\end{equation}
Now, expanding the logarithmic function with respect to $(\delta\Delta^-)^2$, we can further simplify it by keeping only the leading term,
\begin{equation}
\begin{split}
S_{\text{eff}}\simeq&\frac{1}{2}\sum_q\delta \Delta^-_q\left[\frac{2}{u_6-u^{(1)}_4}+2\Pi_q\right.\\
&\left.+\sum_{\alpha}\frac{\lambda}{r_0+\gamma|\nu_n|+f_{\alpha,\mathbf{q}}}\right]\delta \Delta^-_{-q}.\end{split}
\end{equation}
As the temperature approaches $T_N$ from above, the first two terms in the coefficient of this Gaussian term remain finite and positive, but the last term is negative and diverges. Therefore, the coefficient approaches zero and eventually turns into the negative regime. This signals a new instability, setting in prior to the divergence of the magnetic susceptibility. This instability occurs in the direct channels and it is similar to Pomeranchuk instabilities.\cite{Pomeranchuk} This implies a deformation of the FS in the ground state.\cite{doi:10.1146/annurev-conmatphys-070909-103925} Also note that since $\Pi_q>0$, this instability cannot arise without the aid of magnetic fluctuations, and nesting between electron and hole pockets is essential in this scenario. 

\section{Nematic ordered phase}\label{MF}

So far, we have shown that the isotropic metallic phase can be unstable near $T_N$ due to magnetic fluctuations. This suggests that a new phase occurs between the isotropic metallic and the magnetic phase. In this section, we will identify this as nematic ordering. The effective action for the nematic order parameter will be derived within mean field theory.

\subsection{Mean field theory}
To obtain the Ginzburg-Landau action, one cannot simply replace $\delta\Delta^-$ by its order parameter in \eqref{Seff}. This method works only when the order parameter is small and all the terms beyond the Gaussian approximation are positive. Although the order parameter is small in our case, the quartic term (magnetic-nematic fluctuation coupling) is negative. Any truncation of the series expansion beyond the quartic terms in \eqref{Expansion1} will potentially lead to an ill defined (diverging) functional integral. Therefore, instead of expanding around $\langle\phi_q\rangle=0$ in the $\phi$-field manifold, we only expand $\Delta^+$, $M^X$ and $M^Y$ around zero and keep $\Delta^-$ in an arbitrary functional form. Thus, $\mathcal{G}^{-1}_0$ is replaced by $\mathcal{G'}^{-1}_0$,
\begin{equation}
\mathcal{G'}^{-1}_{0,kk'}=\mathcal{G}^{-1}_{0,kk'}+\begin{pmatrix}\Delta^-_{k-k'}& 0 & 0 \\ 0 &-\Delta^-_{k-k'}& 0 \\
0 &  0 & 0 \end{pmatrix},
\end{equation}
and $\mathcal{V}$ is replaced by $\mathcal{V'}$,
\begin{equation}
\mathcal{V'}_{kk'}=\begin{pmatrix}0 & 0 & \delta M^X_{k-k'}
\\0 & 0  & \delta M^Y_{k-k'} \\
\delta M^X_{k-k'} &  \delta M^Y_{k-k'} & 0 \end{pmatrix}.
\end{equation}
Again, the fluctuations of $\Delta^+$ are ignored. Thus, 
\begin{equation}\label{SfullN}
\begin{split}
S_{\text{eff}}&=\frac{1}{2}\sum_q\left[\frac{2}{u_\text{n}}\Delta^-_q\Delta^-_{-q}+\frac{2}{u_\text{spin}}\sum_\alpha M^\alpha_q M^\alpha_{-q}\right]\\
&-\text{tr}\text{ln}\mathcal{G'}_0^{-1}-\text{tr}\text{ln}[1+\mathcal{G'}_0\mathcal{V'}].
\end{split}
\end{equation}

So far, the above equations have not assumed any functional form for $\Delta^-_q$. In order to proceed, we assume that the stable ground state energetically favors homogeneous configurations in space (preserving translational symmetry) and time (static). Then, we can invert $\mathcal{G'}^{-1}_0$ exactly and expand \eqref{SfullN} around $\langle \Delta^+_q \rangle = \langle M^\alpha_q\rangle=0$
\begin{equation}\label{SNSpin}
S_{\text{eff}}\simeq S_{\text{nem}}[\Delta^-]+\frac{1}{2}\sum_{\alpha,q}\delta M^\alpha_q\tilde{\chi}^{-1}_{\alpha,q}\delta M^\alpha_{-q},
\end{equation}
where
\begin{equation}\label{SN}
\begin{split}
S_{\text{nem}}&[\Delta^-]=\frac{1}{2}\frac{2}{u_\text{n}}(\Delta^-)^2\\
&-\sum_k\text{ln}[G^{-1}_{X,k}+\Delta^-][G^{-1}_{Y,k}-\Delta^-],
\end{split}
\end{equation}
and $\tilde{\chi}^{-1}_{\alpha,q}=2/u_\text{spin}+2\sum_k 1/[G^{-1}_{\Gamma,k} (G^{-1}_{\alpha,k+q}\pm\Delta^-)]$, where the upper (lower) sign refers to $\alpha=X$ ($\alpha=Y$). The effective action in \eqref{SNSpin} is our main result that describes the low-energy magnetic collective modes in the nematic phase. $S_{\text{nem}}$ is the Ginzburg-Landau action of the nematic order parameter, and the second term describes the magnetic fluctuations. The original magnetic and nematic fluctuations coupling term in \eqref{Seff} is implicitly contained in the second term of Eq. \eqref{SNSpin}. One can quickly check that in this phase the magnetic fluctuations are unequal in the $X$ and $Y$ direction. Namely, the susceptibility (the correlation of magnetic fluctuations) is $\langle\delta M^\alpha_k \delta M^\alpha_{k'} \rangle\sim\tilde{\chi}_{\alpha,k-k'}$, which is $\Delta^-$ dependent.

Furthermore, at this point the system still possesses a SDW instability, since $\tilde{\chi}^{-1}_{\alpha,q}$ approaches zero at a finite temperature $T_N$. But $T_N$ may be different from the prediction in isotropic phase, because non-zero values of $\Delta^-$ modify the nesting condition between the electron and hole FS pocket. This modification lifts the degeneracies between stripe orders.

Note that the magnetic fluctuations peak around  $\mathbf{Q}_1$ and $\mathbf{Q}_2$ or $q=0$ at low energies. Therefore, it is natural to impose a cutoff frequency $\omega_c$ for $\delta M^\alpha_q$. This frequency is also the characteristic frequency that describes the typical wavelength of the magnetic fluctuations. To make a connection with the phenomenology of magnetic susceptibilities, we can constrain ourself to the fluctuating modes below this cutoff. Hence again, by expanding $\tilde{\chi}^{-1}_{\alpha,q}$ around $q=0$, we obtain $\tilde{\chi}^{-1}_{\alpha,q}=\tilde{\chi}^{-1}_{\alpha,q=0}+\gamma|\nu_n|+f_{\alpha,\mathbf{q}}$, but $\gamma$ and $f_{\alpha,\mathbf{q}}$ can be different in the isotropic phase. If the FS renormalization is not too drastic in nematic phase, we can further expand $\tilde{\chi}^{-1}_{\alpha,q=0}$ around the isotropic phase. Thus, we have $\tilde{\chi}^{-1}_{\alpha,q=0}=r_0+\kappa\Delta^-$, where $\kappa$ is some constant that depends on the microscopic details of the material. This result coincides with Ref. \onlinecite{PhysRevLett.107.217002}.

The mean field equation of the nematic order parameter can be obtained by varying $S_{\text{nem}}$ with respect to $\Delta^-$,
\begin{equation}
\frac{2\Delta^-}{u_\text{n}}=-\sum_k\left[\frac{1}{G^{-1}_{X,k}+\Delta^-}-\frac{1}{G^{-1}_{Y,k}-\Delta^-}\right].
\end{equation}
This equation can be straightforwardly be solved numerically, giving the location of the minimum of $S_{\text{nem}}$. One can also quickly check that $\Delta^-=0$ satisfies the mean field equation trivially.  

\subsection{Nematic ordering phase transition}
To search for the non-zero of $\Delta^-$, instead of evaluating the mean field equation, we plot the value of $S_{\text{nem}}[\Delta^-]$ versus $\Delta^-$ at a given temperature numerically. This  demonstrate how $S_{\text{nem}}$ can develop a minimum at non-zero $\Delta^-$, as temperature is varied. In order to do this, we evaluate the Matsubara sum in \eqref{SN},
\begin{equation}
\begin{split}
&S_{\text{nem}}[\Delta^-]=\frac{1}{2}\frac{2}{u_\text{n}}(\Delta^-)^2\\
&-\sum_\mathbf{k}\text{ln}[1+e^{-\beta(\epsilon^{X}_{\mathbf{k}}+\Delta^-)}][1+e^{-\beta(\epsilon^{Y}_{\mathbf{k}}-\Delta^-)}].
\end{split}
\end{equation}
As shown in Sec. \ref{Instabilities}, the isotropic state is unstable only when the magnetic fluctuations become important. Therefore, it is natural to take $\omega_c$ as the cutoff energy for this effective model. Then, $\sum_\mathbf{k}\rightarrow\int_0^{\omega_c}\frac{m}{(2\pi)^2}d\epsilon\int_0^{2\pi}d\theta$. 

\begin{figure}[htbp]
   \centering
   \includegraphics[scale=0.8]{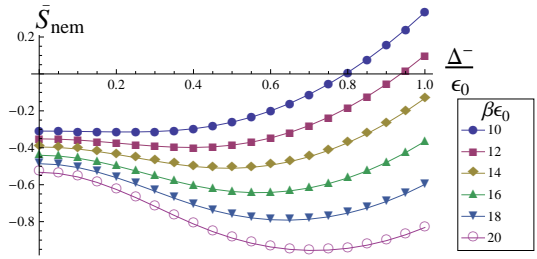}
   \caption{The nematic order phase transition occurs at some finite temperature where $\bar{S}_\text{nem}=S_\text{nem}(2\pi)^2/(m\epsilon_0^2)$.}
   \label{Nem}
\end{figure}

The input parameters are $\delta_2\simeq0.2\epsilon_0$, $\mu=0.05\epsilon_0$,\cite{PhysRevLett.107.217002} where $\epsilon_0\simeq 0.2\text{ eV}$,\cite{1367-2630-11-2-025016} and the magnetic fluctuations cutoff energy $\omega_c=3.4\text{ meV}$.\cite{PhysRevLett.105.157003} In addition, we set $a=(1/u_n)(2\pi)^2(\epsilon_0/m)$, which is a dimensionless parameter controlling the transition temperature in the mean field approximation. For demonstration purposes, we set $u_n=0.1$ and use the hole Fermi pocket momentum from Ref. \onlinecite{1367-2630-11-2-025016}. Thus we have $a\simeq1.5$. Using these input parameters, we obtain the plot shownthe plot is in Fig. \ref{Nem}.

According to Fig. \ref{Nem}, the transition temperature from the mean field approximation is $\beta\epsilon_0\simeq12$ or $T\simeq193\text{ K}$ and $\Delta^-\simeq0.4\epsilon_0$. These values are clearly overestimates compared to experiments. This is because mean field theory does not taking account the correlations between fluctuations. This deviation is even more pronounced because in the low effective dimensionality. However, the mean field result at least estimates the correct order of magnitude. 

So far, magnetism does not appear to play any direct role in the nematic phase. However, the order parameter depends on the cutoff $\omega_c$ (this is similar to the mean field analysis in BCS theory). This determines the position of the minimum, where by a smaller cutoff implies smaller $\Delta^-$ or lower transition temperature and vice versa. To further investigate the relationship between magnetic fluctuations and nematic order calls for more sophisticated methods beyond mean field theory. 


\section{Discussion}\label{discussions}

\subsection{The control parameter}
For the nematic order considered in this paper, $u_n=u_6-u^{(1)}_4$ is the key parameter. It not only determines whether the system can sustain nematic ordering, but also determines the transition temperature to the nematic order from the isotropic phase. These coupling constants depend on microscopic details, and are difficult to obtain in general. Nevertheless, this could explain the fact that nematic ordering is not found in every FeAs material. 

Why $u_\text{n}$ plays such a crucial role in this scenario can be understood by the following intuitive picture. $u_6$ and $u^{(1)}_4$ are related to the coupling strength between two quasi-particles, depending on whether both reside in different electron FS pockets or the same electron FS pocket respectively. If $u_6 > u^{(1)}_4$ means that two quasi-particles tend to occupy the same electron pocket rather than different pockets because this yields a smaller potential energy. Therefore, instabilities of the original isotropic ground state may arise, whenever this scattering processes surpass the other channels.

\subsection{Comparison with other work}

The nematic order considered in this study also yields an anisotropic magnetic fluctuation, which can explain results from recent neutron scattering experiment.\cite{PhysRevB.81.214407} It also implies anisotropic resistivity in the nematic phase.\cite{Chu13082010,PhysRevB.81.184508,PhysRevLett.107.217002} 

In earlier theoretical work, it was shown explicitly that nematic order can be developed solely from the effect of magnetic fluctuations, and yields an anisotropic magnetic fluctuating state. In the phenomenological approach based on a $J_1-J_2$ model,\cite{PhysRevLett.101.076401,PhysRevB.77.224509,PhysRevB.78.020501} or a preemptive nematic order from a microscopic itinerant model,\cite{PhysRevB.85.024534} nematicity originates from the competition between two different stripe orders. Another alternative explanation was given by orbital ordering.\cite{PhysRevB.82.045125} In this scenario, the interactions between orbitals drive local ordering in orbital occupancy, and resulting in rotational symmetry breaking. However, all of these consequences are  similar to our approach. It is the FS pocket in $X$ and $Y$ that are renormalized unequally in the ground state. 

In our approach, the nematic order is interpreted using a quasi-particle picture. Because of the instability in the direct channel, the FS fluctuations (the excitations of the quasi-particles) can yield lower energies than its original isotropic FS configuration. Therefore the system tends to deform an isotropic FS to lower the ground state energy. In addition, only the Gaussian terms of the magnetic fluctuation were kept in our study. Hence, the competition with stripe orders does not play an essential role in this scenario. However, it would be interesting to study the interplay between stripes and nematic order ($\Delta^-$) by including fluctuations beyond the Gaussian approximation. Furthermore, the work in this paper is based on an itinerant picture. How the local ordering relates to the approach of this work is another interesting topic to be explored in the future.

\subsection{Concluding remarks}
In summary, we have discussed possible instabilities of the isotropic metallic phase within an itinerant model. By using the Hubbard-Stratonovich transformation, several auxiliary bosonic fields that describe the magnetic and nematic order collective modes were introduced. By perturbing the action of these bosonic fields in the isotropic phase, if the inter electron pocket interaction ($u_6$) is greater than the intra electron pocket interaction ($u^{(1)}_4$), the magnetic fluctuations can drive an instability in the direct channels.  From mean field analysis, we argue that  nematic order is a favored ground state which spontaneously breaks $C_4$ symmetry but preserves translational symmetry. In the nematic ordered state, the size of the FS pockets in $X$ and $Y$ becomes unequal. For $\Delta^->0$ ($\Delta^-<0$), the X (Y) pocket is larger than the Y (X) pocket. Because of this distortion, the magnetic fluctuations also exhibit anisotropy in X and Y directions. A similar nematic order can possibly occur in FS hole pocket, if angular depend in the coupling constants is considered. 

\acknowledgements
This work is supported by the Department of Energy under Grant No. DE-FG02-05ER46240.

\appendix
\section{Hubburb-Stratonovich Transformation}
The identity of the `fat' unity of the Gaussian functional integral of $\phi$-field is given by
\begin{equation}\label{Unity}
\begin{split}
\mathbf{1}=&\int\mathcal{D}[\phi]\exp\left[-\frac{1}{2}\sum_{\alpha,q}\frac{2}{u_\text{spin}}\mathbf{M}^\alpha_q\cdot \mathbf{M}^\alpha_{-q}\right.\\
&\left.-\frac{1}{2}\sum_{q}\left(\frac{2}{u'_\text{n}}\Delta^+_q\Delta^+_{-q}+\frac{2}{u_\text{n}}\Delta^-_q\Delta^-_{-q}\right)\right].
\end{split}
\end{equation}
Here, $u_\text{n}$ must be positive to ensure the integral in \eqref{Unity} is convergent, and the $\mathbf{1}$ (`fat' unity) represents the normalization constant. 

The quartic fermonic interaction terms in \eqref{Unity} can be generated by shifting the auxiliary fields by 
\begin{equation}\label{shift1}
\mathbf{M}^{\alpha}_q\rightarrow \mathbf{M}^{\alpha}_q+\frac{u_s}{4}\mathbf{s}^\alpha_{q},
\end{equation}
\begin{equation}\label{shift2}
\Delta^{+}_q\rightarrow \Delta^+_q+i\frac{u'_\text{n}}{4}[\rho^X_{q}+\rho^Y_{q}],
\end{equation}
\begin{equation}\label{shift3}
\Delta^{-}_q\rightarrow \Delta^-_q+\frac{u_\text{n}}{4}[\rho^X_{q}-\rho^Y_{q}],
\end{equation}
where $\mathbf{s}^{\alpha}_q= \sum_k\bar{\psi}^\Gamma_{k+q,s}\bm{\sigma}_{ss'}\psi^\alpha_{ks'}$, $\rho^\alpha_{q}=\sum_{k}\bar{\psi}^\alpha_{k+q,s}\psi^\alpha_{ks}$, and $k=(i\omega_n,\mathbf{k})$. $\omega_n$ is the Fermionic Matsubara frequency. Since $u'_{\text{n}}>0$, the shift in $\Delta^+$ includes an extra $i$ factor. This is ensures that the generated quartic fermonic interaction terms carry opposite signs and cancel with those in \eqref{Hint}.

\bibliography{FeAsBib}

\begin{thebibliography}{32}%
\makeatletter
\providecommand \@ifxundefined [1]{%
 \@ifx{#1\undefined}
}%
\providecommand \@ifnum [1]{%
 \ifnum #1\expandafter \@firstoftwo
 \else \expandafter \@secondoftwo
 \fi
}%
\providecommand \@ifx [1]{%
 \ifx #1\expandafter \@firstoftwo
 \else \expandafter \@secondoftwo
 \fi
}%
\providecommand \natexlab [1]{#1}%
\providecommand \enquote  [1]{``#1''}%
\providecommand \bibnamefont  [1]{#1}%
\providecommand \bibfnamefont [1]{#1}%
\providecommand \citenamefont [1]{#1}%
\providecommand \href@noop [0]{\@secondoftwo}%
\providecommand \href [0]{\begingroup \@sanitize@url \@href}%
\providecommand \@href[1]{\@@startlink{#1}\@@href}%
\providecommand \@@href[1]{\endgroup#1\@@endlink}%
\providecommand \@sanitize@url [0]{\catcode `\\12\catcode `\$12\catcode
  `\&12\catcode `\#12\catcode `\^12\catcode `\_12\catcode `\%12\relax}%
\providecommand \@@startlink[1]{}%
\providecommand \@@endlink[0]{}%
\providecommand \url  [0]{\begingroup\@sanitize@url \@url }%
\providecommand \@url [1]{\endgroup\@href {#1}{\urlprefix }}%
\providecommand \urlprefix  [0]{URL }%
\providecommand \Eprint [0]{\href }%
\providecommand \doibase [0]{http://dx.doi.org/}%
\providecommand \selectlanguage [0]{\@gobble}%
\providecommand \bibinfo  [0]{\@secondoftwo}%
\providecommand \bibfield  [0]{\@secondoftwo}%
\providecommand \translation [1]{[#1]}%
\providecommand \BibitemOpen [0]{}%
\providecommand \bibitemStop [0]{}%
\providecommand \bibitemNoStop [0]{.\EOS\space}%
\providecommand \EOS [0]{\spacefactor3000\relax}%
\providecommand \BibitemShut  [1]{\csname bibitem#1\endcsname}%
\let\auto@bib@innerbib\@empty
\bibitem [{\citenamefont {Kamihara}\ \emph {et~al.}(2008)\citenamefont
  {Kamihara}, \citenamefont {Watanabe}, \citenamefont {Hirano},\ and\
  \citenamefont {Hosono}}]{doi:10.1021/ja800073m}%
  \BibitemOpen
  \bibfield  {author} {\bibinfo {author} {\bibfnamefont {Y.}~\bibnamefont
  {Kamihara}}, \bibinfo {author} {\bibfnamefont {T.}~\bibnamefont {Watanabe}},
  \bibinfo {author} {\bibfnamefont {M.}~\bibnamefont {Hirano}}, \ and\ \bibinfo
  {author} {\bibfnamefont {H.}~\bibnamefont {Hosono}},\ }\href {\doibase
  10.1021/ja800073m} {\bibfield  {journal} {\bibinfo  {journal} {Journal of the
  American Chemical Society}\ }\textbf {\bibinfo {volume} {130}},\ \bibinfo
  {pages} {3296} (\bibinfo {year} {2008})}\BibitemShut {NoStop}%
\bibitem [{\citenamefont {Stewart}(2011)}]{Stewart:2011fk}%
  \BibitemOpen
  \bibfield  {author} {\bibinfo {author} {\bibfnamefont {G.~R.}\ \bibnamefont
  {Stewart}},\ }\href {\doibase 10.1103/RevModPhys.83.1589} {\bibfield
  {journal} {\bibinfo  {journal} {Rev. Mod. Phys.}\ }\textbf {\bibinfo {volume}
  {83}},\ \bibinfo {pages} {1589} (\bibinfo {year} {2011})}\BibitemShut
  {NoStop}%
\bibitem [{\citenamefont {Johnston}(2010)}]{johnston2010puzzle}%
  \BibitemOpen
  \bibfield  {author} {\bibinfo {author} {\bibfnamefont {D.~C.}\ \bibnamefont
  {Johnston}},\ }\href@noop {} {\bibfield  {journal} {\bibinfo  {journal}
  {Advances in Physics}\ }\textbf {\bibinfo {volume} {59}},\ \bibinfo {pages}
  {803} (\bibinfo {year} {2010})}\BibitemShut {NoStop}%
\bibitem [{\citenamefont {Leb\`egue}(2007)}]{PhysRevB.75.035110}%
  \BibitemOpen
  \bibfield  {author} {\bibinfo {author} {\bibfnamefont {S.}~\bibnamefont
  {Leb\`egue}},\ }\href {\doibase 10.1103/PhysRevB.75.035110} {\bibfield
  {journal} {\bibinfo  {journal} {Phys. Rev. B}\ }\textbf {\bibinfo {volume}
  {75}},\ \bibinfo {pages} {035110} (\bibinfo {year} {2007})}\BibitemShut
  {NoStop}%
\bibitem [{\citenamefont {Singh}\ and\ \citenamefont
  {Du}(2008)}]{PhysRevLett.100.237003}%
  \BibitemOpen
  \bibfield  {author} {\bibinfo {author} {\bibfnamefont {D.~J.}\ \bibnamefont
  {Singh}}\ and\ \bibinfo {author} {\bibfnamefont {M.-H.}\ \bibnamefont {Du}},\
  }\href {\doibase 10.1103/PhysRevLett.100.237003} {\bibfield  {journal}
  {\bibinfo  {journal} {Phys. Rev. Lett.}\ }\textbf {\bibinfo {volume} {100}},\
  \bibinfo {pages} {237003} (\bibinfo {year} {2008})}\BibitemShut {NoStop}%
\bibitem [{\citenamefont {Cao}\ \emph {et~al.}(2008)\citenamefont {Cao},
  \citenamefont {Hirschfeld},\ and\ \citenamefont
  {Cheng}}]{PhysRevB.77.220506}%
  \BibitemOpen
  \bibfield  {author} {\bibinfo {author} {\bibfnamefont {C.}~\bibnamefont
  {Cao}}, \bibinfo {author} {\bibfnamefont {P.~J.}\ \bibnamefont {Hirschfeld}},
  \ and\ \bibinfo {author} {\bibfnamefont {H.-P.}\ \bibnamefont {Cheng}},\
  }\href {\doibase 10.1103/PhysRevB.77.220506} {\bibfield  {journal} {\bibinfo
  {journal} {Phys. Rev. B}\ }\textbf {\bibinfo {volume} {77}},\ \bibinfo
  {pages} {220506} (\bibinfo {year} {2008})}\BibitemShut {NoStop}%
\bibitem [{\citenamefont {Graser}\ \emph {et~al.}(2009)\citenamefont {Graser},
  \citenamefont {Maier}, \citenamefont {Hirschfeld},\ and\ \citenamefont
  {Scalapino}}]{1367-2630-11-2-025016}%
  \BibitemOpen
  \bibfield  {author} {\bibinfo {author} {\bibfnamefont {S.}~\bibnamefont
  {Graser}}, \bibinfo {author} {\bibfnamefont {T.~A.}\ \bibnamefont {Maier}},
  \bibinfo {author} {\bibfnamefont {P.~J.}\ \bibnamefont {Hirschfeld}}, \ and\
  \bibinfo {author} {\bibfnamefont {D.~J.}\ \bibnamefont {Scalapino}},\ }\href
  {http://stacks.iop.org/1367-2630/11/i=2/a=025016} {\bibfield  {journal}
  {\bibinfo  {journal} {New Journal of Physics}\ }\textbf {\bibinfo {volume}
  {11}},\ \bibinfo {pages} {025016} (\bibinfo {year} {2009})}\BibitemShut
  {NoStop}%
\bibitem [{\citenamefont
  {Chubukov}(2012)}]{doi:10.1146/annurev-conmatphys-020911-125055}%
  \BibitemOpen
  \bibfield  {author} {\bibinfo {author} {\bibfnamefont {A.}~\bibnamefont
  {Chubukov}},\ }\href@noop {} {\bibfield  {journal} {\bibinfo  {journal}
  {Annual Review of Condensed Matter Physics}\ }\textbf {\bibinfo {volume}
  {3}},\ \bibinfo {pages} {57} (\bibinfo {year} {2012})}\BibitemShut {NoStop}%
\bibitem [{\citenamefont {Chu}\ \emph {et~al.}(2010)\citenamefont {Chu},
  \citenamefont {Analytis}, \citenamefont {De~Greve}, \citenamefont {McMahon},
  \citenamefont {Islam}, \citenamefont {Yamamoto},\ and\ \citenamefont
  {Fisher}}]{Chu13082010}%
  \BibitemOpen
  \bibfield  {author} {\bibinfo {author} {\bibfnamefont {J.-H.}\ \bibnamefont
  {Chu}}, \bibinfo {author} {\bibfnamefont {J.~G.}\ \bibnamefont {Analytis}},
  \bibinfo {author} {\bibfnamefont {K.}~\bibnamefont {De~Greve}}, \bibinfo
  {author} {\bibfnamefont {P.~L.}\ \bibnamefont {McMahon}}, \bibinfo {author}
  {\bibfnamefont {Z.}~\bibnamefont {Islam}}, \bibinfo {author} {\bibfnamefont
  {Y.}~\bibnamefont {Yamamoto}}, \ and\ \bibinfo {author} {\bibfnamefont
  {I.~R.}\ \bibnamefont {Fisher}},\ }\href {\doibase 10.1126/science.1190482}
  {\bibfield  {journal} {\bibinfo  {journal} {Science}\ }\textbf {\bibinfo
  {volume} {329}},\ \bibinfo {pages} {824} (\bibinfo {year}
  {2010})}\BibitemShut {NoStop}%
\bibitem [{\citenamefont {Chu}\ \emph {et~al.}(2012)\citenamefont {Chu},
  \citenamefont {Kuo}, \citenamefont {Analytis},\ and\ \citenamefont
  {Fisher}}]{Chu10082012}%
  \BibitemOpen
  \bibfield  {author} {\bibinfo {author} {\bibfnamefont {J.-H.}\ \bibnamefont
  {Chu}}, \bibinfo {author} {\bibfnamefont {H.-H.}\ \bibnamefont {Kuo}},
  \bibinfo {author} {\bibfnamefont {J.~G.}\ \bibnamefont {Analytis}}, \ and\
  \bibinfo {author} {\bibfnamefont {I.~R.}\ \bibnamefont {Fisher}},\ }\href
  {\doibase 10.1126/science.1221713} {\bibfield  {journal} {\bibinfo  {journal}
  {Science}\ }\textbf {\bibinfo {volume} {337}},\ \bibinfo {pages} {710}
  (\bibinfo {year} {2012})}\BibitemShut {NoStop}%
\bibitem [{\citenamefont {Dusza}\ \emph {et~al.}(2011)\citenamefont {Dusza},
  \citenamefont {Lucarelli}, \citenamefont {Pfuner}, \citenamefont {Chu},
  \citenamefont {Fisher},\ and\ \citenamefont
  {Degiorgi}}]{0295-5075-93-3-37002}%
  \BibitemOpen
  \bibfield  {author} {\bibinfo {author} {\bibfnamefont {A.}~\bibnamefont
  {Dusza}}, \bibinfo {author} {\bibfnamefont {A.}~\bibnamefont {Lucarelli}},
  \bibinfo {author} {\bibfnamefont {F.}~\bibnamefont {Pfuner}}, \bibinfo
  {author} {\bibfnamefont {J.-H.}\ \bibnamefont {Chu}}, \bibinfo {author}
  {\bibfnamefont {I.~R.}\ \bibnamefont {Fisher}}, \ and\ \bibinfo {author}
  {\bibfnamefont {L.}~\bibnamefont {Degiorgi}},\ }\href
  {http://stacks.iop.org/0295-5075/93/i=3/a=37002} {\bibfield  {journal}
  {\bibinfo  {journal} {EPL (Europhysics Letters)}\ }\textbf {\bibinfo {volume}
  {93}},\ \bibinfo {pages} {37002} (\bibinfo {year} {2011})}\BibitemShut
  {NoStop}%
\bibitem [{\citenamefont {Tanatar}\ \emph {et~al.}(2010)\citenamefont
  {Tanatar}, \citenamefont {Blomberg}, \citenamefont {Kreyssig}, \citenamefont
  {Kim}, \citenamefont {Ni}, \citenamefont {Thaler}, \citenamefont {Bud'ko},
  \citenamefont {Canfield}, \citenamefont {Goldman}, \citenamefont {Mazin},\
  and\ \citenamefont {Prozorov}}]{PhysRevB.81.184508}%
  \BibitemOpen
  \bibfield  {author} {\bibinfo {author} {\bibfnamefont {M.~A.}\ \bibnamefont
  {Tanatar}}, \bibinfo {author} {\bibfnamefont {E.~C.}\ \bibnamefont
  {Blomberg}}, \bibinfo {author} {\bibfnamefont {A.}~\bibnamefont {Kreyssig}},
  \bibinfo {author} {\bibfnamefont {M.~G.}\ \bibnamefont {Kim}}, \bibinfo
  {author} {\bibfnamefont {N.}~\bibnamefont {Ni}}, \bibinfo {author}
  {\bibfnamefont {A.}~\bibnamefont {Thaler}}, \bibinfo {author} {\bibfnamefont
  {S.~L.}\ \bibnamefont {Bud'ko}}, \bibinfo {author} {\bibfnamefont {P.~C.}\
  \bibnamefont {Canfield}}, \bibinfo {author} {\bibfnamefont {A.~I.}\
  \bibnamefont {Goldman}}, \bibinfo {author} {\bibfnamefont {I.~I.}\
  \bibnamefont {Mazin}}, \ and\ \bibinfo {author} {\bibfnamefont
  {R.}~\bibnamefont {Prozorov}},\ }\href {\doibase 10.1103/PhysRevB.81.184508}
  {\bibfield  {journal} {\bibinfo  {journal} {Phys. Rev. B}\ }\textbf {\bibinfo
  {volume} {81}},\ \bibinfo {pages} {184508} (\bibinfo {year}
  {2010})}\BibitemShut {NoStop}%
\bibitem [{\citenamefont {Chuang}\ \emph {et~al.}(2010)\citenamefont {Chuang},
  \citenamefont {Allan}, \citenamefont {Lee}, \citenamefont {Xie},
  \citenamefont {Ni}, \citenamefont {Bud'ko}, \citenamefont {Boebinger},
  \citenamefont {Canfield},\ and\ \citenamefont {Davis}}]{Chuang08012010}%
  \BibitemOpen
  \bibfield  {author} {\bibinfo {author} {\bibfnamefont {T.-M.}\ \bibnamefont
  {Chuang}}, \bibinfo {author} {\bibfnamefont {M.~P.}\ \bibnamefont {Allan}},
  \bibinfo {author} {\bibfnamefont {J.}~\bibnamefont {Lee}}, \bibinfo {author}
  {\bibfnamefont {Y.}~\bibnamefont {Xie}}, \bibinfo {author} {\bibfnamefont
  {N.}~\bibnamefont {Ni}}, \bibinfo {author} {\bibfnamefont {S.~L.}\
  \bibnamefont {Bud'ko}}, \bibinfo {author} {\bibfnamefont {G.~S.}\
  \bibnamefont {Boebinger}}, \bibinfo {author} {\bibfnamefont {P.~C.}\
  \bibnamefont {Canfield}}, \ and\ \bibinfo {author} {\bibfnamefont {J.~C.}\
  \bibnamefont {Davis}},\ }\href {\doibase 10.1126/science.1181083} {\bibfield
  {journal} {\bibinfo  {journal} {Science}\ }\textbf {\bibinfo {volume}
  {327}},\ \bibinfo {pages} {181} (\bibinfo {year} {2010})}\BibitemShut
  {NoStop}%
\bibitem [{\citenamefont {Nakajima}\ \emph {et~al.}(2011)\citenamefont
  {Nakajima}, \citenamefont {Liang}, \citenamefont {Ishida}, \citenamefont
  {Tomioka}, \citenamefont {Kihou}, \citenamefont {Lee}, \citenamefont {Iyo},
  \citenamefont {Eisaki}, \citenamefont {Kakeshita}, \citenamefont {Ito},\ and\
  \citenamefont {Uchida}}]{Nakajima26072011}%
  \BibitemOpen
  \bibfield  {author} {\bibinfo {author} {\bibfnamefont {M.}~\bibnamefont
  {Nakajima}}, \bibinfo {author} {\bibfnamefont {T.}~\bibnamefont {Liang}},
  \bibinfo {author} {\bibfnamefont {S.}~\bibnamefont {Ishida}}, \bibinfo
  {author} {\bibfnamefont {Y.}~\bibnamefont {Tomioka}}, \bibinfo {author}
  {\bibfnamefont {K.}~\bibnamefont {Kihou}}, \bibinfo {author} {\bibfnamefont
  {C.~H.}\ \bibnamefont {Lee}}, \bibinfo {author} {\bibfnamefont
  {A.}~\bibnamefont {Iyo}}, \bibinfo {author} {\bibfnamefont {H.}~\bibnamefont
  {Eisaki}}, \bibinfo {author} {\bibfnamefont {T.}~\bibnamefont {Kakeshita}},
  \bibinfo {author} {\bibfnamefont {T.}~\bibnamefont {Ito}}, \ and\ \bibinfo
  {author} {\bibfnamefont {S.}~\bibnamefont {Uchida}},\ }\href {\doibase
  10.1073/pnas.1100102108} {\bibfield  {journal} {\bibinfo  {journal}
  {Proceedings of the National Academy of Sciences}\ }\textbf {\bibinfo
  {volume} {108}},\ \bibinfo {pages} {12238} (\bibinfo {year}
  {2011})}\BibitemShut {NoStop}%
\bibitem [{\citenamefont {Kasahara}\ \emph {et~al.}(2012)\citenamefont
  {Kasahara}, \citenamefont {Shi}, \citenamefont {Hashimoto}, \citenamefont
  {Tonegawa}, \citenamefont {Mizukami}, \citenamefont {Shibauchi},
  \citenamefont {Sugimoto}, \citenamefont {Fukuda}, \citenamefont {Terashima},
  \citenamefont {Nevidomskyy},\ and\ \citenamefont
  {Matsuda}}]{Kasahara:2012fk}%
  \BibitemOpen
  \bibfield  {author} {\bibinfo {author} {\bibfnamefont {S.}~\bibnamefont
  {Kasahara}}, \bibinfo {author} {\bibfnamefont {H.~J.}\ \bibnamefont {Shi}},
  \bibinfo {author} {\bibfnamefont {K.}~\bibnamefont {Hashimoto}}, \bibinfo
  {author} {\bibfnamefont {S.}~\bibnamefont {Tonegawa}}, \bibinfo {author}
  {\bibfnamefont {Y.}~\bibnamefont {Mizukami}}, \bibinfo {author}
  {\bibfnamefont {T.}~\bibnamefont {Shibauchi}}, \bibinfo {author}
  {\bibfnamefont {K.}~\bibnamefont {Sugimoto}}, \bibinfo {author}
  {\bibfnamefont {T.}~\bibnamefont {Fukuda}}, \bibinfo {author} {\bibfnamefont
  {T.}~\bibnamefont {Terashima}}, \bibinfo {author} {\bibfnamefont {A.~H.}\
  \bibnamefont {Nevidomskyy}}, \ and\ \bibinfo {author} {\bibfnamefont
  {Y.}~\bibnamefont {Matsuda}},\ }\href {http://dx.doi.org/10.1038/nature11178}
  {\bibfield  {journal} {\bibinfo  {journal} {Nature}\ }\textbf {\bibinfo
  {volume} {486}},\ \bibinfo {pages} {382} (\bibinfo {year}
  {2012})}\BibitemShut {NoStop}%
\bibitem [{\citenamefont {Fradkin}\ \emph {et~al.}(2010)\citenamefont
  {Fradkin}, \citenamefont {Kivelson}, \citenamefont {Lawler}, \citenamefont
  {Eisenstein},\ and\ \citenamefont
  {Mackenzie}}]{doi:10.1146/annurev-conmatphys-070909-103925}%
  \BibitemOpen
  \bibfield  {author} {\bibinfo {author} {\bibfnamefont {E.}~\bibnamefont
  {Fradkin}}, \bibinfo {author} {\bibfnamefont {S.~A.}\ \bibnamefont
  {Kivelson}}, \bibinfo {author} {\bibfnamefont {M.~J.}\ \bibnamefont
  {Lawler}}, \bibinfo {author} {\bibfnamefont {J.~P.}\ \bibnamefont
  {Eisenstein}}, \ and\ \bibinfo {author} {\bibfnamefont {A.~P.}\ \bibnamefont
  {Mackenzie}},\ }\href {\doibase 10.1146/annurev-conmatphys-070909-103925}
  {\bibfield  {journal} {\bibinfo  {journal} {Annual Review of Condensed Matter
  Physics}\ }\textbf {\bibinfo {volume} {1}},\ \bibinfo {pages} {153} (\bibinfo
  {year} {2010})}\BibitemShut {NoStop}%
\bibitem [{\citenamefont {Si}\ and\ \citenamefont
  {Abrahams}(2008)}]{PhysRevLett.101.076401}%
  \BibitemOpen
  \bibfield  {author} {\bibinfo {author} {\bibfnamefont {Q.}~\bibnamefont
  {Si}}\ and\ \bibinfo {author} {\bibfnamefont {E.}~\bibnamefont {Abrahams}},\
  }\href {\doibase 10.1103/PhysRevLett.101.076401} {\bibfield  {journal}
  {\bibinfo  {journal} {Phys. Rev. Lett.}\ }\textbf {\bibinfo {volume} {101}},\
  \bibinfo {pages} {076401} (\bibinfo {year} {2008})}\BibitemShut {NoStop}%
\bibitem [{\citenamefont {Fang}\ \emph {et~al.}(2008)\citenamefont {Fang},
  \citenamefont {Yao}, \citenamefont {Tsai}, \citenamefont {Hu},\ and\
  \citenamefont {Kivelson}}]{PhysRevB.77.224509}%
  \BibitemOpen
  \bibfield  {author} {\bibinfo {author} {\bibfnamefont {C.}~\bibnamefont
  {Fang}}, \bibinfo {author} {\bibfnamefont {H.}~\bibnamefont {Yao}}, \bibinfo
  {author} {\bibfnamefont {W.-F.}\ \bibnamefont {Tsai}}, \bibinfo {author}
  {\bibfnamefont {J.}~\bibnamefont {Hu}}, \ and\ \bibinfo {author}
  {\bibfnamefont {S.~A.}\ \bibnamefont {Kivelson}},\ }\href {\doibase
  10.1103/PhysRevB.77.224509} {\bibfield  {journal} {\bibinfo  {journal} {Phys.
  Rev. B}\ }\textbf {\bibinfo {volume} {77}},\ \bibinfo {pages} {224509}
  (\bibinfo {year} {2008})}\BibitemShut {NoStop}%
\bibitem [{\citenamefont {Xu}\ \emph {et~al.}(2008)\citenamefont {Xu},
  \citenamefont {M\"uller},\ and\ \citenamefont
  {Sachdev}}]{PhysRevB.78.020501}%
  \BibitemOpen
  \bibfield  {author} {\bibinfo {author} {\bibfnamefont {C.}~\bibnamefont
  {Xu}}, \bibinfo {author} {\bibfnamefont {M.}~\bibnamefont {M\"uller}}, \ and\
  \bibinfo {author} {\bibfnamefont {S.}~\bibnamefont {Sachdev}},\ }\href
  {\doibase 10.1103/PhysRevB.78.020501} {\bibfield  {journal} {\bibinfo
  {journal} {Phys. Rev. B}\ }\textbf {\bibinfo {volume} {78}},\ \bibinfo
  {pages} {020501} (\bibinfo {year} {2008})}\BibitemShut {NoStop}%
\bibitem [{\citenamefont {Fernandes}\ \emph {et~al.}(2012)\citenamefont
  {Fernandes}, \citenamefont {Chubukov}, \citenamefont {Knolle}, \citenamefont
  {Eremin},\ and\ \citenamefont {Schmalian}}]{PhysRevB.85.024534}%
  \BibitemOpen
  \bibfield  {author} {\bibinfo {author} {\bibfnamefont {R.~M.}\ \bibnamefont
  {Fernandes}}, \bibinfo {author} {\bibfnamefont {A.~V.}\ \bibnamefont
  {Chubukov}}, \bibinfo {author} {\bibfnamefont {J.}~\bibnamefont {Knolle}},
  \bibinfo {author} {\bibfnamefont {I.}~\bibnamefont {Eremin}}, \ and\ \bibinfo
  {author} {\bibfnamefont {J.}~\bibnamefont {Schmalian}},\ }\href {\doibase
  10.1103/PhysRevB.85.024534} {\bibfield  {journal} {\bibinfo  {journal} {Phys.
  Rev. B}\ }\textbf {\bibinfo {volume} {85}},\ \bibinfo {pages} {024534}
  (\bibinfo {year} {2012})}\BibitemShut {NoStop}%
\bibitem [{\citenamefont {Zhai}\ \emph {et~al.}(2009)\citenamefont {Zhai},
  \citenamefont {Wang},\ and\ \citenamefont {Lee}}]{PhysRevB.80.064517}%
  \BibitemOpen
  \bibfield  {author} {\bibinfo {author} {\bibfnamefont {H.}~\bibnamefont
  {Zhai}}, \bibinfo {author} {\bibfnamefont {F.}~\bibnamefont {Wang}}, \ and\
  \bibinfo {author} {\bibfnamefont {D.-H.}\ \bibnamefont {Lee}},\ }\href
  {\doibase 10.1103/PhysRevB.80.064517} {\bibfield  {journal} {\bibinfo
  {journal} {Phys. Rev. B}\ }\textbf {\bibinfo {volume} {80}},\ \bibinfo
  {pages} {064517} (\bibinfo {year} {2009})}\BibitemShut {NoStop}%
\bibitem [{\citenamefont {Thomale}\ \emph {et~al.}(2011)\citenamefont
  {Thomale}, \citenamefont {Platt}, \citenamefont {Hanke},\ and\ \citenamefont
  {Bernevig}}]{PhysRevLett.106.187003}%
  \BibitemOpen
  \bibfield  {author} {\bibinfo {author} {\bibfnamefont {R.}~\bibnamefont
  {Thomale}}, \bibinfo {author} {\bibfnamefont {C.}~\bibnamefont {Platt}},
  \bibinfo {author} {\bibfnamefont {W.}~\bibnamefont {Hanke}}, \ and\ \bibinfo
  {author} {\bibfnamefont {B.~A.}\ \bibnamefont {Bernevig}},\ }\href {\doibase
  10.1103/PhysRevLett.106.187003} {\bibfield  {journal} {\bibinfo  {journal}
  {Phys. Rev. Lett.}\ }\textbf {\bibinfo {volume} {106}},\ \bibinfo {pages}
  {187003} (\bibinfo {year} {2011})}\BibitemShut {NoStop}%
\bibitem [{\citenamefont {Altalnd}\ and\ \citenamefont
  {Simons}(2010)}]{SimonQFT}%
  \BibitemOpen
  \bibfield  {author} {\bibinfo {author} {\bibfnamefont {A.}~\bibnamefont
  {Altalnd}}\ and\ \bibinfo {author} {\bibfnamefont {B.~D.}\ \bibnamefont
  {Simons}},\ }\href@noop {} {\emph {\bibinfo {title} {Condensed Matter Field
  Theory}}},\ \bibinfo {edition} {2nd}\ ed.\ (\bibinfo  {publisher} {Cambrige
  University Press},\ \bibinfo {year} {2010})\BibitemShut {NoStop}%
\bibitem [{\citenamefont {Maiti}\ and\ \citenamefont
  {Chubukov}(2010)}]{PhysRevB.82.214515}%
  \BibitemOpen
  \bibfield  {author} {\bibinfo {author} {\bibfnamefont {S.}~\bibnamefont
  {Maiti}}\ and\ \bibinfo {author} {\bibfnamefont {A.~V.}\ \bibnamefont
  {Chubukov}},\ }\href {\doibase 10.1103/PhysRevB.82.214515} {\bibfield
  {journal} {\bibinfo  {journal} {Phys. Rev. B}\ }\textbf {\bibinfo {volume}
  {82}},\ \bibinfo {pages} {214515} (\bibinfo {year} {2010})}\BibitemShut
  {NoStop}%
\bibitem [{\citenamefont {Luttinger}(1960)}]{PhysRev.119.1153}%
  \BibitemOpen
  \bibfield  {author} {\bibinfo {author} {\bibfnamefont {J.~M.}\ \bibnamefont
  {Luttinger}},\ }\href {\doibase 10.1103/PhysRev.119.1153} {\bibfield
  {journal} {\bibinfo  {journal} {Phys. Rev.}\ }\textbf {\bibinfo {volume}
  {119}},\ \bibinfo {pages} {1153} (\bibinfo {year} {1960})}\BibitemShut
  {NoStop}%
\bibitem [{\citenamefont {Metzner}\ \emph {et~al.}(2012)\citenamefont
  {Metzner}, \citenamefont {Salmhofer}, \citenamefont {Honerkamp},
  \citenamefont {Meden},\ and\ \citenamefont
  {Sch\"onhammer}}]{RevModPhys.84.299}%
  \BibitemOpen
  \bibfield  {author} {\bibinfo {author} {\bibfnamefont {W.}~\bibnamefont
  {Metzner}}, \bibinfo {author} {\bibfnamefont {M.}~\bibnamefont {Salmhofer}},
  \bibinfo {author} {\bibfnamefont {C.}~\bibnamefont {Honerkamp}}, \bibinfo
  {author} {\bibfnamefont {V.}~\bibnamefont {Meden}}, \ and\ \bibinfo {author}
  {\bibfnamefont {K.}~\bibnamefont {Sch\"onhammer}},\ }\href {\doibase
  10.1103/RevModPhys.84.299} {\bibfield  {journal} {\bibinfo  {journal} {Rev.
  Mod. Phys.}\ }\textbf {\bibinfo {volume} {84}},\ \bibinfo {pages} {299}
  (\bibinfo {year} {2012})}\BibitemShut {NoStop}%
\bibitem [{\citenamefont {Hertz}(1976)}]{PhysRevB.14.1165}%
  \BibitemOpen
  \bibfield  {author} {\bibinfo {author} {\bibfnamefont {J.~A.}\ \bibnamefont
  {Hertz}},\ }\href {\doibase 10.1103/PhysRevB.14.1165} {\bibfield  {journal}
  {\bibinfo  {journal} {Phys. Rev. B}\ }\textbf {\bibinfo {volume} {14}},\
  \bibinfo {pages} {1165} (\bibinfo {year} {1976})}\BibitemShut {NoStop}%
\bibitem [{\citenamefont {Pomeranchuk}(1958)}]{Pomeranchuk}%
  \BibitemOpen
  \bibfield  {author} {\bibinfo {author} {\bibfnamefont {I.~J.}\ \bibnamefont
  {Pomeranchuk}},\ }\href@noop {} {\bibfield  {journal} {\bibinfo  {journal}
  {Sov. Phys. JTEP}\ }\textbf {\bibinfo {volume} {8}},\ \bibinfo {pages} {361}
  (\bibinfo {year} {1958})}\BibitemShut {NoStop}%
\bibitem [{\citenamefont {Fernandes}\ \emph {et~al.}(2011)\citenamefont
  {Fernandes}, \citenamefont {Abrahams},\ and\ \citenamefont
  {Schmalian}}]{PhysRevLett.107.217002}%
  \BibitemOpen
  \bibfield  {author} {\bibinfo {author} {\bibfnamefont {R.~M.}\ \bibnamefont
  {Fernandes}}, \bibinfo {author} {\bibfnamefont {E.}~\bibnamefont {Abrahams}},
  \ and\ \bibinfo {author} {\bibfnamefont {J.}~\bibnamefont {Schmalian}},\
  }\href {\doibase 10.1103/PhysRevLett.107.217002} {\bibfield  {journal}
  {\bibinfo  {journal} {Phys. Rev. Lett.}\ }\textbf {\bibinfo {volume} {107}},\
  \bibinfo {pages} {217002} (\bibinfo {year} {2011})}\BibitemShut {NoStop}%
\bibitem [{\citenamefont {Fernandes}\ \emph {et~al.}(2010)\citenamefont
  {Fernandes}, \citenamefont {VanBebber}, \citenamefont {Bhattacharya},
  \citenamefont {Chandra}, \citenamefont {Keppens}, \citenamefont {Mandrus},
  \citenamefont {McGuire}, \citenamefont {Sales}, \citenamefont {Sefat},\ and\
  \citenamefont {Schmalian}}]{PhysRevLett.105.157003}%
  \BibitemOpen
  \bibfield  {author} {\bibinfo {author} {\bibfnamefont {R.~M.}\ \bibnamefont
  {Fernandes}}, \bibinfo {author} {\bibfnamefont {L.~H.}\ \bibnamefont
  {VanBebber}}, \bibinfo {author} {\bibfnamefont {S.}~\bibnamefont
  {Bhattacharya}}, \bibinfo {author} {\bibfnamefont {P.}~\bibnamefont
  {Chandra}}, \bibinfo {author} {\bibfnamefont {V.}~\bibnamefont {Keppens}},
  \bibinfo {author} {\bibfnamefont {D.}~\bibnamefont {Mandrus}}, \bibinfo
  {author} {\bibfnamefont {M.~A.}\ \bibnamefont {McGuire}}, \bibinfo {author}
  {\bibfnamefont {B.~C.}\ \bibnamefont {Sales}}, \bibinfo {author}
  {\bibfnamefont {A.~S.}\ \bibnamefont {Sefat}}, \ and\ \bibinfo {author}
  {\bibfnamefont {J.}~\bibnamefont {Schmalian}},\ }\href {\doibase
  10.1103/PhysRevLett.105.157003} {\bibfield  {journal} {\bibinfo  {journal}
  {Phys. Rev. Lett.}\ }\textbf {\bibinfo {volume} {105}},\ \bibinfo {pages}
  {157003} (\bibinfo {year} {2010})}\BibitemShut {NoStop}%
\bibitem [{\citenamefont {Diallo}\ \emph {et~al.}(2010)\citenamefont {Diallo},
  \citenamefont {Pratt}, \citenamefont {Fernandes}, \citenamefont {Tian},
  \citenamefont {Zarestky}, \citenamefont {Lumsden}, \citenamefont {Perring},
  \citenamefont {Broholm}, \citenamefont {Ni}, \citenamefont {Bud'ko},
  \citenamefont {Canfield}, \citenamefont {Li}, \citenamefont {Vaknin},
  \citenamefont {Kreyssig}, \citenamefont {Goldman},\ and\ \citenamefont
  {McQueeney}}]{PhysRevB.81.214407}%
  \BibitemOpen
  \bibfield  {author} {\bibinfo {author} {\bibfnamefont {S.~O.}\ \bibnamefont
  {Diallo}}, \bibinfo {author} {\bibfnamefont {D.~K.}\ \bibnamefont {Pratt}},
  \bibinfo {author} {\bibfnamefont {R.~M.}\ \bibnamefont {Fernandes}}, \bibinfo
  {author} {\bibfnamefont {W.}~\bibnamefont {Tian}}, \bibinfo {author}
  {\bibfnamefont {J.~L.}\ \bibnamefont {Zarestky}}, \bibinfo {author}
  {\bibfnamefont {M.}~\bibnamefont {Lumsden}}, \bibinfo {author} {\bibfnamefont
  {T.~G.}\ \bibnamefont {Perring}}, \bibinfo {author} {\bibfnamefont {C.~L.}\
  \bibnamefont {Broholm}}, \bibinfo {author} {\bibfnamefont {N.}~\bibnamefont
  {Ni}}, \bibinfo {author} {\bibfnamefont {S.~L.}\ \bibnamefont {Bud'ko}},
  \bibinfo {author} {\bibfnamefont {P.~C.}\ \bibnamefont {Canfield}}, \bibinfo
  {author} {\bibfnamefont {H.-F.}\ \bibnamefont {Li}}, \bibinfo {author}
  {\bibfnamefont {D.}~\bibnamefont {Vaknin}}, \bibinfo {author} {\bibfnamefont
  {A.}~\bibnamefont {Kreyssig}}, \bibinfo {author} {\bibfnamefont {A.~I.}\
  \bibnamefont {Goldman}}, \ and\ \bibinfo {author} {\bibfnamefont {R.~J.}\
  \bibnamefont {McQueeney}},\ }\href {\doibase 10.1103/PhysRevB.81.214407}
  {\bibfield  {journal} {\bibinfo  {journal} {Phys. Rev. B}\ }\textbf {\bibinfo
  {volume} {81}},\ \bibinfo {pages} {214407} (\bibinfo {year}
  {2010})}\BibitemShut {NoStop}%
\bibitem [{\citenamefont {Lv}\ \emph {et~al.}(2010)\citenamefont {Lv},
  \citenamefont {Kr\"uger},\ and\ \citenamefont
  {Phillips}}]{PhysRevB.82.045125}%
  \BibitemOpen
  \bibfield  {author} {\bibinfo {author} {\bibfnamefont {W.}~\bibnamefont
  {Lv}}, \bibinfo {author} {\bibfnamefont {F.}~\bibnamefont {Kr\"uger}}, \ and\
  \bibinfo {author} {\bibfnamefont {P.}~\bibnamefont {Phillips}},\ }\href
  {\doibase 10.1103/PhysRevB.82.045125} {\bibfield  {journal} {\bibinfo
  {journal} {Phys. Rev. B}\ }\textbf {\bibinfo {volume} {82}},\ \bibinfo
  {pages} {045125} (\bibinfo {year} {2010})}\BibitemShut {NoStop}%
\end{thebibliography}%

\end{document}